# Migration of Near Earth Asteroid to Jovian-Crosser Asteroid :
# Case Study 3552 Don Quixote

*Suryadi Siregar*
*Bosscha Observatory and Astronomy Research Division,*
*Faculty of Mathematics and Natural Sciences, Institut Teknologi Bandung*
*e-mail: suryadi@as.itb.ac.id*

*Abstract*

It is generally recognized that main-belt asteroids (MBAs) and nuclei of extinct comets are the two main sources for the Near-Earth-Asteroids (NEAs). Theoretical studies of NEAs dynamics and numerical modelling of their orbital motions showed that the resonance mechanism for supplying NEAs is quite sufficient to sustain this population.
Asteroid 1983 SA, also known as 3552 Don Quixote, is one of Near Earth Asteroids (NEAs) and the most probable candidates for NEAs of the cometary origin. In this work, an investigation on the evolution of the orbit is done by using the SWIFT subroutine package, where the gravitational perturbations of eight planets: Mercury, Venus, Earth, Mars, Jupiter, Saturn, Uranus, and Neptune are considered. Migration of asteroid 3552 Don Quixote from Near Earth Asteroids (NEAs) to Jovian-crosser asteroid is found.

*Abstrak*

Asteroid sabuk utama (MBAs) dan komet yang telah mati umumnya dianggap sebagai dua sumber utama asteroid yang melintas dekat Bumi (NEAs). Telaah teoritis tentang dinamika NEA dan model numerik gerak orbitnya menunjukkan bahwa mekanisme resonansi sebagai pemasok utama NEAs sudah sangat cukup untuk mempertahankan populasi ini.
Asteroid 1983 SA dikenal juga sebagai 3552 Don Quixote adalah satu dari asteroid dekat Bumi (NEAs) dan yang paling mungkin berasal dari komet. Dalam pekerjaan ini penelitian evolusi orbit dilakukan dengan menggunakan paket subroutine SWIFT, dan dengan memperhitungkan gangguan gravitasi delapan planet: Merkurius, Venus, Bumi, Mars, Jupiter, Saturnus dan Neptunus. Terdapat perpindahan asteroid 3552 Don Quixote dari Near Earth Asteroids (NEAs) ke Jovian-crosser asteroid.

*Kata kunci*: Asteroid, Element orbit, Tata surya, Masalah N-benda

## 1. Introduction

Asteroid 3552 Don Quixote is the most probable candidate for NEAs of the cometary origin. Other objects originating from comet are 2100 Ra-Shalom, 2101 Adonis, 2201 Oljato, 2212 Hephaistos, 3200 Phaeton, and 4015 Wilson-Harrington (Lupishko and Martino, 2000). All the NEAs have small size, almost the same variety within their taxonomic classes and mineralogy, with a predominance of differentiated assemblages among them, approximately similar shape and rotation, optical properties and surface structure, compared to those of MBAs. All these evidences clearly indicate that the main belt asteroid is the principal source of NEAs, while comet nuclei contribution to the total NEAs population does not exceed 10% (Yoshida and Nakamura, 2004). By definition NEAs are approaching asteroids which orbit the Earth. Qualitatively defined as asteroids with perihelion distance smaller than 1.3 AU. Otherwise asteroids with orbits intersecting trajectories Jovian planets (Jupiter, Saturn, Uranus, and Neptune) are known as Jovian-crosser asteroids. The groups of NEAs are presented in Table 1.

Investigation of orbital evolution of 3552 Don Quixote had already been done by taking into account all the planets in the Solar System as perturbing bodies including Pluto, with the integration-time up to 7500 years (Dirgantari, 1999). On August 24, 2006 the International Astronomical Union (IAU) defined the term of planet. A planet is a celestial body that (resolution B5):
1. is in orbit of the Sun
2. has sufficient mass to assume hydrostatic equilibrium (a nearly round shape)
3. has cleared the neighbourhood around its orbit

This definition excludes Pluto as a planet. In view of the IAU resolution it is necessary to review the status of orbital evolution of 3552 Don Quixote. Perturbation of small bodies in the Solar System is relatively small compared to perturbation of planets. This study only concerned about perturbation from planets.

**Table 1.** Description and orbital criteria of NEAs groups

| No | Group | Description | Definition |
|---|---|---|---|
| 1 | Atiras | NEAs whose orbits are inside the orbit of the Earth (named after asteroid 163693 Atira) | $Q < 0.983$ AU |
| 2 | Atens | Earth-crossing NEAs with semimajor axes smaller than Earth's orbit (named after asteroid 2062 Aten) | $a < 1.0$ AU, $Q > 0.983$ AU |
| 3 | Apollos | Earth-crossing NEAs with semimajor axes larger than Earth's orbit (named after asteroid 1862 Apollo) | $a > 1.0$ AU, $q < 1.017$ AU |
| 4 | Amors | Earth-approaching NEAs with orbits between Earth's orbit and Mars' orbit (named after asteroid 1221 Amor) | $a > 1.0$ AU, $1.017 < q < 1.3$ AU |

Note $a$ = semimajor axis, $Q$ = aphelion distance, $q$ = perihelion distance

To obtain a comprehensive overview of the orbital evolution of 3552 Don Quixote, a numerical integration of the Solar System motion was taken over 250000 years. In this range of time, our Sun is stable and still on the main-sequence in the Herztprung-Russel Diagram. This consideration implies that orbit of asteroid will not be affected by radiation from the Sun. In gigayear time interval, asteroid's evolutionary model and observational results are often inconsistent, especially for sub-km-sized asteroids. The SWIFT software package still can be used and enhanced by incorporating thermal radiation effects due to evolutionary effect in the Sun (see for example, Fermita and Dermawan, 2010 ).

Another example the application of SWIFT package has been conducted by Roig and Ferraz-Mello (1999) to investigate the origin of Zongguo-type asteroids and Griqua-like objects.

## 2. Orbital elements and physical parameters

By using the observation data from 1983 to 2009, the orbital elements of 3522 Don Quixote are already catalogued. Table 2 and Table 3 present all the information about the orbital elements and physical characteristics of 3552 Don Quixote respectively, taken from http://ssd.jplnasa.gov.

**Table 2.** Orbital elements of 3552 Don Quixote as test particle. All the elements refer to epoch MJD 24455200.5 (2010-Jan 04.0)

| Element | Symbol | Value | Units |
|---|---|---|---|
| Eccentricity | $e$ | 0.7138803024084498 | |
| Semimajor axis | $a$ | 4.224923970817345 | AU |
| Perihelion distance | $q$ | 1.20883396887755 | AU |
| inclination | $i$ | 30.96930692119038 | deg |
| Ascending node | $\Omega$ | 350.2673100000013 | deg |
| Perihelion argument | $\omega$ | 317.1044852074207 | deg |
| Mean anomaly at epoch | $M$ | 13.32372560374069 | deg |
| Perihelion passage | $T_o$ | 2455083.104904571105 (2009-Sep-08.6049046) | JED |
| Sideral period | $P$ | 3171.95322174279 8.68 | day yr |
| Daily motion | $n$ | 0.113494738047304 | deg/day |
| Aphelion distance | $Q$ | 7.24101397275714 | AU |

**Table 3.** Physical characteristics of 3552 Don Quixote

| Parameter | Symbol | Value | Units |
|---|---|---|---|
| Absolute magnitude | H | 13.0 | mag |
| Diameter | D | 19.0 | km |
| Rotation Period | Prot | 7.7 | hour |
| Geometric Albedo | Albedo | 0.03 | |

## 3. The N-body problem

Let a system of N-bodies consists of point masses $M_i$ at $r_i$ where $i=1, 2, .., n$. Consider the equation of motion of a body of negligible mass, i.e. an asteroid in solar system. If the origin is taken at the centre of the Sun and $r$ is the position vector of the asteroid, we have the acceleration vector of asteroid motion given by;

$$r = -GM_\Theta \frac{r}{|r|^3} \sum_{i=1}^{n} GM_i \left( \frac{r - r_i}{|r - r_i|^3} + \frac{r_i}{|r_i|^3} \right) \quad (1)$$

where: $G$- gravitational constant, $M_\Theta$ - mass of the Sun, $M_i$ - mass of the $i^{th}$ –planet.

The first term represents acceleration force due to the Sun, the second and the third ones acceleration forces by planets to asteroid and by planet to the Sun, respectively.
Illustration of this system is given in Figure 1.

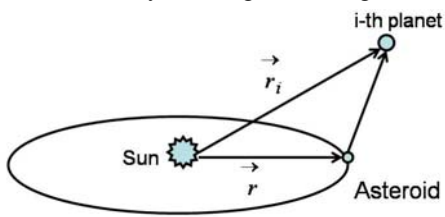

**Figure 1.** Gravitational force influences from the Sun and planet on asteroid

To solve this equation, the computer program SWIFT created by Duncan and Levison (http://www.boulder.swri.edu) is used. The SWIFT package provided here is designed to integrate a set of mutually gravitationally interacting bodies together with a group of test particle under the gravitational influence of the massive bodies but do not affect the massive bodies. In this work Regularized Mixed Variable Symplectic (RMVS) method is used. This is an extension of the Wisdom-Holman Mapping (WHM) that handles close approaches between test particles and planets (Levison and Duncan, 1994).

For the initial conditions, we used the orbital data of principal planets in our solar system issued by NASA (http://ssd.jplnasa.gov). These data are presented in Table 4.

**Table 4.** Orbital elements and masses of eight planets as perturbing bodies. Epoch MJD 24455200.5 (2010-Jan 04.0)

| No | Planet | $a$ [AU] | $e$ [-] | $i$ [deg] | $M$ [deg] | $\Omega$ [deg] | $\omega$ [deg] | $M$ [M$_\odot$] |
|---|---|---|---|---|---|---|---|---|
| 1 | Mercury | 0.387099 | 0.205634 | 7.004 | 174.794 | 48.331 | 77.455 | 1/6023600.0 |
| 2 | Venus | 0.723332 | 0.006773 | 3.394 | 50.407 | 76.680 | 131.571 | 1/408523.5 |
| 3 | Earth | 1.000000 | 0.016709 | 0.000 | 357.525 | 174.876 | 102.940 | 1/328900.5 |
| 4 | Mars | 1.523692 | 0.093405 | 1.849 | 19.387 | 49.557 | 336.059 | 1/3098710.0 |
| 5 | Jupiter | 5.202437 | 0.048402 | 1.304 | 250.327 | 100.468 | 15.719 | 1/1047.355 |
| 6 | Saturn | 9.551712 | 0.052340 | 2.485 | 267.246 | 113.643 | 90.968 | 1/3498.5 |
| 7 | Uranus | 19.293108 | 0.044846 | 0.773 | 118.432 | 74.090 | 176.615 | 1/22869.0 |
| 8 | Neptune | 30.257162 | 0.007985 | 1.770 | 292.471 | 131.775 | 3.096 | 1/19314.0 |

## 4. Results and Discussions

Figure 2 shows the evolution of the semimajor axes of the planets over 250000 years. Epoch 0 means the present situation. We see that planets' semimajor axes are relatively stable up to 230000 years.

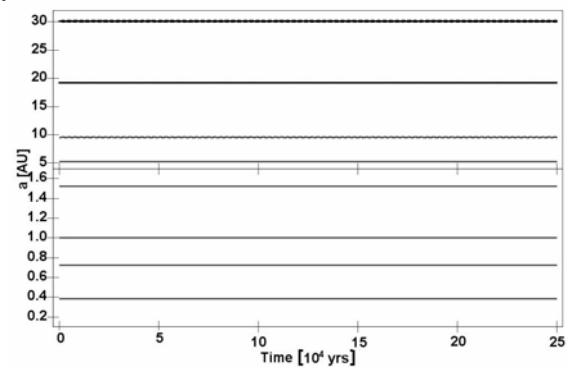

**Figure 2.** Graph of the simulated evolution of semimajor axes of planets up to 250,000 years. The curves from bottom to top represent Mercury, Venus, Earth, Mars, Jupiter, Saturn, Uranus, and Neptune, respectively.

Figure 2 also shows that over 250000 years the distances of planets to the Sun are relatively constant. The orbits of planets are essentially not affected by the Solar System's small variation gravitation. These situations are completely different from the semimajor axis of 3552 Don Quixote, where orbit's eccentricity changes rapidly. The orbit of 3552 Don Quixote is not stable and chaotic as shown in Figure 3.

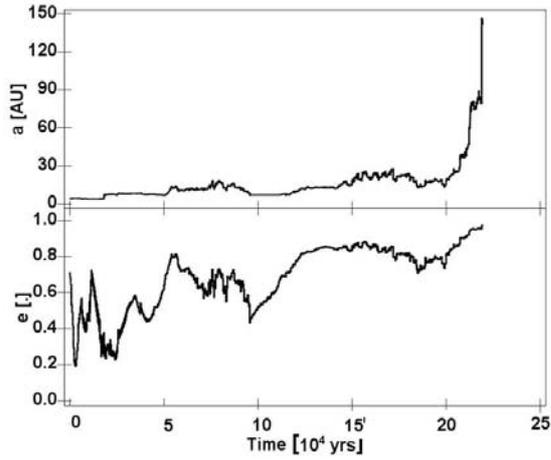

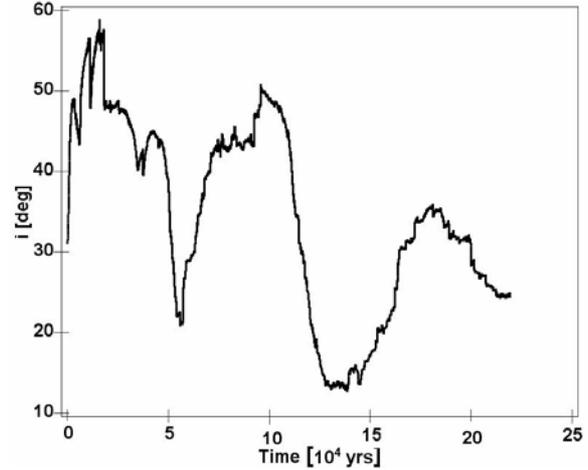

**Figure 3.** Graph of the simulated evolution of semimajor axis and eccentricity of 3552 Don Quixote. In the extreme right, the semimajor axis and eccentricity monotonically increase. Singular point is reached at t~ 219301 years where the gravitational influences of planets are not significant anymore. The asteroid probably escapes from our solar system.

Based on Figure 3, consider the maximum in abscissa, $t_{max}$ ~ 219301 years from now, where semimajor axis and eccentricity are 143.2 AU and 0.9737, respectively. According to Newtonian orbit, these values conclude that the period of revolution $P \sim a^{3/2}$ is ~ 1713.62 yr, or approximately 197.4 times the present period. It means the asteroid will move more slowly than they do now. Its perihelion distance q~3.7 AU, and aphelion distance, Q~282.6 AU imply that the escape velocity from our solar system at aphelion point is $V = \sqrt{\frac{2GM}{Q}}$ = 2.5 km s$^{-1}$. This value is very small compared to Earth's escape velocity which is ~11.2 km s$^{-1}$. If we consider that the dominant perturbation on 3552 Don Quixote originates from Jupiter, according to three-body restricted problem, Tisserand invariant of the Sun-Jupiter-Asteroid system is T~2.184, which informs that the gravitation of the Solar System does not contribute at all. These quantities suggest that 3552 Don Quixote does not sustain its orbit in the Solar System. By definition of asteroid distance to the Sun, $q < r < Q$, it is clear that 3552 Don Quixote is an asteroid whose orbit crosses that of Jovian (Jupiter, Saturn, Uranus and Neptune). Hence occurs a migration of asteroid 3552 Don Quixote from Near Earth Asteroids (NEAs) to Jovian-crosser asteroid.

**Figure 4.** Evolution of inclination of 3552 Don Quixote, over 219301 years from now.

Inclination over 219301 years present some singularities. This phenomenon means that inclination of 3552 Don Quixote changes rapidly, for examples at time, t~ $10^4$, $2\times10^4$, $6\times10^4$ etc. Consider the Tisserand invariant:

$$T = \frac{1}{a} + 2\sqrt{a(1-e^2)}\cos i \qquad (2)$$

The second term of equation (2), known as Kozai's resonance $H = \sqrt{a(1-e^2)}\cos i$ remains constant since $T$ is always constant. If $a(1 - e^2)$ decreases, $\cos i$ increases. It means that the orbital plane will approach the ecliptic. In the opposite case, if $a(1 - e^2)$ increases, $\cos i$ decreases. It means that the orbital plane recedes from the ecliptic. This conclusion explains the behaviours of semimajor axis, eccentricity, and inclinations of 3552 Don Quixote as shown in Figure 3 and Figure 4. To explain the detailed physical processes of these phenomena further studies are absolutely necessary. Research on the evolution of the orbit of 3552 Don Quixote has been done by Benest et al (1985) but the integration time is taken back to 1000 years counted from the year 1985. The orbital elements of 1000 years ago almost do not differ when compared to those of present time..

**Conclusions**

This investigation demonstrates that from a dynamical point of view, 3552 Don Quixote must be considered as serious candidate that will leave the Solar System. This information will enrich our knowledge about the origin of asteroids, and provide a better understanding about the evolution of the Solar System. In further study we plan to investigate the probability of 3552 Don Quixote escapes from the Solar System.

**Acknowledgment**

The author is indebted to Dr. Budi Dermawan for fruitful discussion and computer programming. A part of this report was already presented in The 5th Kyoto University Southeast Asia Forum, Conference of the Earth and Space Sciences, Bandung 7-8 January 2010.